\documentclass[12pt]{article}
\title{Evaluation of a $\ln \tan$ integral arising in quantum field theory}
\author{Mark W. Coffey\\
Department of Physics\\
Colorado School of Mines\\
Golden, CO  80401\\
(Received $\mbox{~~~~~~~~~~~~~~~~~~~~~~~~~~~~~~~2007}$)}
\date{August 27, 2007}
\begin{document}
\maketitle
\begin{center}

\baselineskip=25 pt
\begin{abstract}

We analytically evaluate a dilogarithmic integral that is prototypical of
volumes of ideal tetrahedra in hyperbolic geometry.  We additionally obtain 
new representations of the Clausen function Cl$_2$ and the Catalan constant $G=\mbox{Cl}_2(\pi/2)$, as well as new relations between sine and Clausen 
function values. 

\end{abstract}

\vfill
\baselineskip=15pt
\centerline{\bf Key words and phrases}
\medskip

\noindent
Clausen function, trigamma function, polygamma function, dilogarithm function, Hurwitz zeta function 
\end{center}

\bigskip
\centerline{\bf AMS classification numbers}
33B30, 33B15, 11M35   


\baselineskip=25pt
\pagebreak
\medskip
\centerline{\bf Introduction and statement of results}
\medskip

In this paper, we provide a closed form evaluation of the integral
$$I_7 \equiv {{24} \over {7 \sqrt{7}}}\int_{\pi/3}^{\pi/2} \ln \left|{{\tan t+\sqrt{7}} \over {\tan t-\sqrt{7}}}\right |dt, \eqno(1.1)$$
in terms of elementary mathematical constants and values of the Clausen
function Cl$_2$.  
This and related integrals originate in hyperbolic geometry and quantum field
theory \cite{broad,lunev}.  The integral $I_7$ arises in the analysis of the
volume of ideal tetrahedra in hyperbolic space and is the simplest of 998
empirically investigated cases wherein the volume of a hyperbolic knot
complement appears to be expressible in terms of a Dirichlet L series
\cite{bb98}.  In particular, Borwein and Broadhurst have conjectured that
$$I_7 \stackrel{?}{=}L_{-7}(2)=\sum_{n=0}^\infty\left [{1 \over {(7n+1)^2}}+
{1 \over {(7n+2)^2}}-{1 \over {(7n+3)^2}}+{1 \over {(7n+4)^2}}-{1 \over {(7n+5)^2}}-{1 \over {(7n+6)^2}}\right].  \eqno(1.2)$$
The ? here indicates numerical verification has been performed but that no proof exists.

The integral (1.1) has been much discussed of late \cite{bbcomp}, and in fact it provides the opening equation of a very recent book \cite{bb07}.  The evaluation 
of $I_7$ in the specific form of Eq. (1.2) remains to be completed--our
evaluation is in terms of a certain other combination of Clausen function values.
In addition to the Clausen value relations conjectured in Ref. \cite{bb98},
our work suggests further, presumably kindred, relations.  Moreover, our
work may provide a means of making progress in the other 
997 cases linking a knot complement volume to an L series.  The equality 
of expressions such as Eqs. (1.1) and (1.2) would tightly link topics in 
hyperbolic geometry with analytic number theory.

In the rest of this Introduction, we present two Lemmas, state our main
result, and provide some background relations on the particular Clausen
function Cl$_2$.

It seems worth while to note some of the numerous alternative expressions
for the value $L_{-7}(2)$ and its companion Clausen value combination.  We have
{\newline \bf Lemma 1}.  Let $\psi'$ be the trigamma function and
$\zeta(s,a)$ the Hurwitz zeta function (e.g, \cite{nbs}).  (a) We have 
$$L_{-7}(2)={1 \over {49}}\left [\psi'\left({1 \over 7}\right)+\psi'\left({2 
\over 7}\right)-\psi'\left({3 \over 7}\right)+\psi'\left({4 \over 7}\right)-\psi'\left({5 \over 7}\right)-\psi'\left({6 \over 7}\right)\right] \eqno(1.3a)$$
$$={1 \over {49}}\left [2\left(\psi'\left({1 \over 7}\right)+\psi'\left({2 
\over 7}\right)-\psi'\left({3 \over 7}\right)\right)-\pi^2\left(\mbox{csc}^2
{\pi \over 7}+ \mbox{csc}^2{{2\pi} \over 7}- \mbox{csc}^2{{3\pi} \over 7}
\right)\right]  \eqno(1.3b)$$ 
$$={2 \over {49}}\left [\psi'\left({1 \over 7}\right)+\psi'\left({2 
\over 7}\right)-\psi'\left({3 \over 7}\right)+\left(\mbox{csc}^2
{{3\pi} \over 7}-4\right)\pi^2 \right]  \eqno(1.3c)$$ 
$$={1 \over {49}}\left [\zeta\left(2,{1 \over 7}\right)+\zeta\left(2,{2 \over 7}\right)-\zeta \left(2,{3 \over 7}\right)+\zeta\left(2,{4 \over 7}\right) -\zeta\left(2,{5 \over 7}\right)-\zeta\left(2,{6 \over 7}\right)\right]   \eqno(1.4)$$
$$=-\int_0^1 {{(1+u-u^2+u^3-u^4-u^5)} \over {(1-u^7)}}
\ln u ~du$$
$$=-\int_0^1 {{(1+2u+u^2+2u^3+u^4)} \over {(1+u+u^2+u^3+u^4+u^5+u^6)}}\ln u ~du$$
$$=1-\int_0^1 {{u(1+u-u^4-u^5)} \over {(1+u+u^2+u^3+u^4+u^5+u^6)}}\ln u ~du
\eqno(1.5)$$
(b) We have
$$\mbox{Cl}_2\left({{2\pi} \over 7}\right)+\mbox{Cl}_2\left({{4\pi} \over 7}
\right)-\mbox{Cl}_2\left({{6\pi} \over 7}\right)={1 \over {56\sqrt{7}}}\left
\{8\left[\psi'\left({1 \over 7}\right)+\psi'\left({2 \over 7}\right)-\psi'\left(
{3 \over 7}\right)\right] \right.$$
$$\left. \pi^2\left(-\mbox{csc}^2 {\pi \over {14}}+2\mbox{csc}^2 {{3\pi} \over {7}}+
+\mbox{csc}^2 {\pi \over {7}}+\mbox{csc}^2 {{3\pi} \over {14}}+\mbox{csc}^2 {{2\pi} \over {7}}+\mbox{csc}^2 {{5\pi} \over {14}}-\mbox{csc}^2 {{4\pi} \over {7}}\right) \right \}.  \eqno(1.6)$$

In Proposition 1 we make use of some of the representations for Cl$_2$
given in the following
{\newline \bf Lemma 2}.  Let $a$ be a real number and set
$$\theta =\cos^{-1}\left({{1-a^2} \over {1+a^2}}\right).  \eqno(1.7)$$
Then we have (a)
$$\mbox{Cl}_2(\theta)=\int_a^\infty \ln\left({{u+a} \over {u-a}}\right)
{{du} \over {1+u^2}}.  \eqno(1.8)$$
Let $H_n$ be the $n$th harmonic number, $\gamma=-\psi(1)$ the Euler constant,
and $\psi$ the digamma function.  Then
(b)
$$\mbox{Cl}_2(\theta)=2 \cot^{-1} a \ln 2+{1 \over a}\sum_{j=1}^\infty {{(-1)^j}
\over a^{2j}} {H_j \over {(2j+1)}} \eqno(1.9a)$$
$$=\cot^{-1} a (2 \ln 2+\gamma-2)+{1 \over a}(2-\gamma)-{1 \over a}\ln\left(1+
{1 \over a^2}\right)+{1 \over a}\sum_{j=1}^\infty {{(-1)^j} \over a^{2j}}
{{\psi(j)} \over {(2j+1)}}.  \eqno(1.9b)$$
(c) We have the integral representation
$$\mbox{Cl}_2(\theta)=2 \cot^{-1} a \ln 2+2a\int_1^\infty {{[y\cot^{-1} ay
-\cot^{-1} a]} \over {(1-y^2)}} {{dy} \over y}.  \eqno(1.10)$$

As an immediate consequence of Lemma 2(b) at $a=1$ we obtain a harmonic
number series representation for the Catalan constant $G=\mbox{Cl}_2(\pi/2)
=1-1/3^2+1/5^2-1/7^2 + \ldots \simeq 0.9159655941$:
{\newline \bf Corollary 1}.  We have
$$G={\pi \over 2}\ln 2+\sum_{j=1}^\infty {{(-1)^j} \over {(2j+1)}} H_j .
\eqno(1.11)$$
This expression has been previously obtained in the very recent Ref.
\cite{borwein07}.

Let $\theta_\pm = \pm \tan^{-1}(\sqrt{7}/3)$,
$$r_{73} \equiv {{\sqrt{7}+\sqrt{3}} \over {\sqrt{7}-\sqrt{3}}}, ~~~~~~~~
\omega_\pm=\tan^{-1}\left({{r_{73} \sin \theta_\pm} \over {1-r_{73}\cos \theta_\pm}}\right), \eqno(1.12a)$$
so that
$$\omega_+=-\cot^{-1} (2\sqrt{3}-\sqrt{7})=\tan^{-1} \sqrt{7}-{{2\pi} \over 3},$$
$$\omega_-=-\omega_+=\tan^{-1}\left({{2\sqrt{3}-\sqrt{7}} \over 5}\right), ~~~~ \mbox{and} ~~~~ \omega_+-\omega_-=-2\omega_+=-2\cot^{-1}(2\sqrt{3}-\sqrt{7}).  \eqno(1.12b)$$
{\bf Proposition 1}.  
We have
$$I_7={{24}\over {7\sqrt{7}}}\left\{\mbox{Cl}_2(\theta_+)+{1 \over 2}[\mbox{Cl}_2(2\omega_+)-\mbox{Cl}_2(2\omega_++2\theta_+)]\right\}.  
\eqno(1.13)$$

The Clausen function Cl$_2$ can be defined by (e.g., \cite{lewin1991})
$$\mbox{Cl}_2(\theta)\equiv -\int_0^\theta\ln\left(2 \sin {t \over 2}\right)dt
=\int_0^1 \tan^{-1}\left({{x \sin \theta} \over {1-x\cos \theta}}\right)
{{dx} \over x}$$
$$=-\sin \theta \int_0^1 {{\ln x} \over {x^2-2x\cos \theta +1}}dx
=\sum_{n=1}^\infty {{\sin(n \theta)} \over n^2}.  \eqno(1.14)$$
When $\theta$ is a rational multiple of $\pi$ it is known that Cl$_2(\theta)$
may be written in terms of the trigamma and sine functions \cite{grosjean,ded}.
As we shall find useful, the imaginary part of the dilogarithm function Li$_2$ of
complex argument may be written in terms of Cl$_2$:
$$\mbox{Im Li}_2(re^{i\theta})=\omega \ln r+{1 \over 2}\left[\mbox{Cl}_2(2\omega)
-\mbox{Cl}_2(2\omega+2\theta) + \mbox{Cl}_2(2\theta)\right],  \eqno(1.15)$$
where
$$\omega=\tan^{-1}\left({{r \sin \theta} \over {1-r\cos \theta}}\right).
\eqno(1.16)$$
Besides being periodic, Cl$_2(\theta)=\mbox{Cl}_2(\theta+2\pi)$, the Clausen
function satisfies the duplication formula
$${1 \over 2}\mbox{Cl}_2(2\theta)=\mbox{Cl}_2(\theta)-\mbox{Cl}_2(\pi-\theta).
\eqno(1.17)$$
Further properties of Cl$_2$ are given in \cite{lewin} and \cite{lewin1991}
and Ref. \cite{prsa03} gives a very recent but necessarily incomplete review
of the dilogarithm function with complex argument.


We next provide the proofs of Lemmas 1 and 2 and Proposition 1 and finish
with a discussion with further analytic results.  Our proof of Proposition 1 
shows how to make a ready generalization of the integral $I_7$.  



\centerline{\bf Proofs of Lemmas 1 and 2, and expressions for the Catalan
constant}
\medskip

For Eqs. (1.3) and (1.4) of Lemma 1, we repeatedly apply the relation
\cite{grad} (p. 944) between the specific Hurwitz zeta function $\zeta(2,a)$
and the trigamma function contained in
$$\sum_{j=0}^\infty {1 \over {(7j+p)^q}}={1 \over 7^q}\zeta\left(q,{p \over 7}
\right)={{(-1)^q} \over 7^q}{1 \over {(q-1)!}}\psi^{(q-1)}\left({p \over 7}
\right),  \eqno(2.1)$$
and use the reflection formula satisfied by the trigamma function
$$\psi'(1-x)=-\psi'(x)+\pi^2 \mbox{csc}^2 \pi x. \eqno(2.2)$$
Equation (1.3c) follows from (1.3b) by the use of the identity
$$\mbox{csc}^2{\pi \over 7}+ \mbox{csc}^2{{2\pi} \over 7}+\mbox{csc}^2{{3\pi} \over 7}=8.  \eqno(2.3)$$ 
For Eq. (1.5), we use the integral representation 
$$\zeta(s,a)={1 \over {\Gamma(s)}}\int_0^\infty {{t^{s-1} e^{-(a-1)t}} \over
{e^t-1}}dt, ~~~~~~\mbox{Re} ~s>1, ~~~\mbox{Re} ~a>0, \eqno(2.4)$$
where $\Gamma$ is the Gamma function, and change variable to $u=e^{-t/7}$.

For part (b) of Lemma 1, we first apply for $p$ even and $q$ odd, with $q \geq 3$
the formula \cite{ded} (p. 329)
$$\mbox{Cl}_2\left({{p \pi} \over q}\right)=-{1 \over {4q^2}}\sum_{k=1}^{q-1}
\left[\psi'\left(1-{k \over {2q}}\right)+\psi'\left({1 \over 2}-{k \over {2q}}
\right)\right] \sin k{p \over q}\pi.  \eqno(2.5)$$
We then obtain
$$\mbox{Cl}_2\left({{2\pi} \over 7}\right)+\mbox{Cl}_2\left({{4\pi} \over 7}
\right)-\mbox{Cl}_2\left({{6\pi} \over 7}\right)$$
$$={1 \over {56\sqrt{7}}}\left
[\psi'\left({1 \over {14}}\right)+\psi'\left({1 \over 7}\right)-\psi'\left(
{3 \over {14}}\right)+\psi'\left({2 \over 7}\right)-\psi'\left({5 \over {14}} \right)-\psi'\left({3 \over 7}\right)\right.$$
$$\left. +\psi'\left({4 \over 7}\right)+\psi'\left({9 \over {14}}\right)-\psi'
\left({5 \over 7}\right)+\psi'\left({{11} \over {14}}\right)-\psi'\left({6 \over 7} \right)-\psi'\left({{13} \over {14}}\right) \right],  \eqno(2.6)$$
wherein we used the identity
$$\sin {{2x \pi} \over 7}+\sin {{4x \pi} \over 7}-\sin {{6x \pi} \over 7}
=\pm {\sqrt{7} \over 2}, \eqno(2.7)$$
for $x=1,\ldots,6$ and the $-$ sign holds on the right side for $x=3, 5$, and $6$.

In regard to Eq. (2.7), we observe that $\pm \sin (2\pi/7)$, $\pm \sin (4\pi/7)$, 
and $\pm \sin (6\pi/7)$ are the nonzero roots of the Chebyshev polynomial
$T_7(x)$.  Indeed, if we write the cubic polynomials
$$p_1(x)=\left(x- \sin {{2 \pi} \over 7}\right)\left(x- \sin {{4 \pi} \over 7}\right)\left(x+ \sin {{6 \pi} \over 7}\right)=x^3-{\sqrt{7} \over 2}x^2+{\sqrt{7}
\over 8}, \eqno(2.8a)$$
and
$$p_2(x)=\left(x- \sin {{6 \pi} \over 7}\right)\left(x+ \sin {{2 \pi} \over 7}\right)\left(x+ \sin {{4 \pi} \over 7}\right)=x^3+{\sqrt{7} \over 2}x^2-{\sqrt{7}
\over 8}, \eqno(2.8b)$$
we then have the factorization $p_1(x)p_2(x)=T_7(x)/64x$.

We next repeatedly apply the duplication formula
$$2\psi'(2x)={1 \over 2}\left[\psi'(x)+\psi'\left(x+{1 \over 2}\right)\right]
\eqno(2.9)$$
and the reflection formula (2.2) in order to reduce the dozen trigamma values
in Eq. (2.6) to only three independent function values.  In particular, we have
the relations
$$\psi'\left({1 \over {14}}\right)=4\psi'\left({1 \over {7}}\right)+
\psi'\left({3 \over {7}}\right)-\pi^2 \mbox{csc}^2 {{4\pi} \over 7}, \eqno(2.10a)$$
$$\psi'\left({3 \over {14}}\right)=4\psi'\left({3 \over {7}}\right)+
\psi'\left({2 \over {7}}\right)-\pi^2 \mbox{csc}^2 {{2\pi} \over 7}, \eqno(2.10b)$$
and
$$\psi'\left({5 \over {14}}\right)=-4\psi'\left({2 \over {7}}\right)+
\psi'\left({1 \over {7}}\right)-\pi^2 \mbox{csc}^2 {\pi \over 7}+4\pi^2 \mbox{csc}^2 {{2\pi} \over 7} \eqno(2.10c)$$
and such substitutions give Eq. (1.6).  

Equation (1.6) could be further 
rewritten with Eq. (2.3) and/or the identity
$$\mbox{csc}^2{\pi \over {14}}+ \mbox{csc}^2{\pi \over 7}+\mbox{csc}^2{{3\pi}
\over {14}}+\mbox{csc}^2{{2\pi} \over 7}+ \mbox{csc}^2{{5\pi} \over {14}} +\mbox{csc}^2{{3\pi} \over 7}=32,  \eqno(2.11)$$
that is another case of the relation $\sum_{j=1}^{[(n-1)/2]} \mbox{csc}^2\left({{j \pi} \over n} \right)=(n^2-1)/6-[1+(-1)^n]/4$ (e.g., \cite{hansen}, p. 260).
Indeed, there are at most two independent trigamma values involved in either
part (a) or (b) of Lemma 1.  As follows from the multiplication formula for the polygamma functions, we have \cite{nbs} (p. 260)
$$\psi'(mx)={1 \over m^2}\sum_{k=0}^{m-1} \psi'\left(x+{k \over m}\right), \eqno(2.12)$$
so that we have the additional relation
$$\zeta(2)=\psi'(1)={1 \over {49}}\sum_{k=0}^6  \psi'\left({{k+1} \over 7}\right).
\eqno(2.13)$$

For part (a) of Lemma 2, we consider the integrals
$$\int_a^\infty \ln^n\left({{u+a} \over {u-a}}\right){{du} \over {1+u^2}}  
={{2a} \over {(1+a^2)}}\int_1^\infty {{\ln^n v ~dv} \over {v^2+2\left({{a^2-1}
\over {a^2+1}}\right)v+1}}, \eqno(2.14)$$
where we changed variable to $v=(u+a)/(u-a)$.  With the further change of
variable $y=1/v$ and use of the definition (1.7) for $\theta$ we have
$$\int_a^\infty \ln^n\left({{u+a} \over {u-a}}\right){{du} \over {1+u^2}}  
={{2a (-1)^n} \over {(1+a^2)}}\int_0^1 {{\ln^n y ~dy} \over {y^2-2\cos \theta ~y
+1}}. \eqno(2.15)$$
Therefore, by the integral representation of Cl$_2$ (1.14), at $n=1$ we obtain
Eq. (1.8).

For part (b), we start with the result of part (a), and change variable to
$y=1/u$ to obtain
$$\mbox{Cl}_2(\theta)=\int_0^{1/a} \ln\left({{1+ay} \over {1-ay}}\right)
{{dy} \over {1+y^2}}.  \eqno(2.16)$$
We then make use of the relation \cite{grad} (p. 1041)
$$\ln\left({{1+z} \over {1-z}}\right)=2z ~_2F_1\left({1 \over 2},1;{3 \over 2}; z^2
\right), \eqno(2.17)$$
where $_2F_1$ is the Gauss hypergeometric function.  Changing variable
again to $w=a^2y^2$ and expanding the factor $(1+y^2)^{-1}$ as a geometric
series we have
$$\mbox{Cl}_2(\theta)={1 \over {2a}}\sum_{j=0}^\infty {{(-1)^j} \over a^{2j}}
\int_0^1 w^j ~_2F_1\left({1 \over 2},1;{3 \over 2}; w\right)dw. \eqno(2.18)$$
Performing the integration term by term, we have
$$\int_0^1 w^j ~_2F_1\left({1 \over 2},1;{3 \over 2}; w\right)dw
=\sum_{\ell=0}^\infty {1 \over {2\ell+1}}\int_0^1 w^{j+\ell}dw$$
$$=\sum_{\ell=0}^\infty {1 \over {(2\ell+1)(j+\ell+1)}}
={1 \over {(2j+1)}}[\psi(j+1)+\gamma+2\ln 2], \eqno(2.19)$$
where the partial fractions form of the digamma function was used.
Since
$$\psi(j+1)+\gamma=H_j, \eqno(2.20)$$
we have found the representation (1.9a).  For Eq. (1.9b) we use the
functional equation of the digamma function, $\psi(j+1)=\psi(j)+1/j$.
The absolute convergence of integrals such as in Eq. (2.9) justifies the
interchange of integration and summation above and throughout this paper.

For Eq. (1.10), we use an integral representation of the digamma function
\cite{grad} (p. 943) to write
$$\sum_{j=1}^\infty {{(-1)^j} \over a^{2j}}{1 \over {2j+1}}[\psi(j+1)+\gamma]
=\sum_{j=1}^\infty {{(-1)^j} \over a^{2j}} {1 \over {2j+1}}\int_0^1 {{(t^j-1)}
\over {(t-1)}}dt$$
$$=a \int_0^1 \left[\cot^{-1}\left({a \over \sqrt{t}}\right){1 \over \sqrt{t}}
-\cot^{-1}a\right]dt, \eqno(2.21)$$
where we interchanged summation and integration.  With the change of variable
$y=1/\sqrt{t}$, we arrive at the integral representation (1.10).

{\bf Remarks} on the Catalan constant.  We have noted that Corollary 1
directly follows from Lemma 2(b).  Another means to find the result (1.11)
is to use the integral relation
$$G={\pi \over 2}\ln 2-{1 \over 2}\int_0^1 {{\ln(1+u)} \over {(1+u)}}{{du} 
\over \sqrt{u}}.  \eqno(2.22)$$
Then the use of the generating function
$$-{{\ln(1+u)} \over {(1+u)}}=\sum_{n=1}^\infty (-1)^n H_n u^n, \eqno(2.23)$$
returns Eq. (1.11).

Another form of the Catalan constant may be found by expanding in the 
integrand of Eq. (2.22) to write
$$G={\pi \over 2}\ln 2-{1 \over 2}\sum_{j=0}^\infty (-1)^j \int_0^1 \ln(1+u)
u^{j-1/2} du.  \eqno(2.24)$$ 
Omitting the details, we determine
$$G=-{\pi \over 4}\ln 2+{1 \over 2}\sum_{j=0}^\infty {{(-1)^j} \over {(2j+1)}}
\left[\psi\left({j \over 2}+{3 \over 4}\right)-\psi\left({j \over 2}+{1 \over 4}\right)\right].  \eqno(2.25)$$
The sum in this equation may be re-expressed by applying the duplication 
formula for the digamma function.

With $\theta=\theta(a)$ defined as in Eq. (1.7), Eq. (1.8) may be rewritten as
$$\mbox{Cl}_2(\theta)=\cot^{-1}a \ln a+2 \int_a^\infty \mbox{coth}^{-1}\left(
{u \over a}\right) {{du} \over {1+u^2}}.  \eqno(2.26)$$
Then we obtain
$$G=2\int_1^\infty {{\mbox{coth}^{-1} u} \over {1+u^2}}du.  \eqno(2.27)$$
We next have
{\newline \bf Lemma 3}.  We have
$$G=-{1 \over 4}\int_0^1 {{\ln x/2} \over {(1-x/2)}}{{dx} \over \sqrt{1-x^2}}
\eqno(2.28a)$$
$$={\pi \over 4}\ln 2+{\sqrt{\pi} \over 4}\sum_{j=0}^\infty {{j!} \over 
{2^j \Gamma(j+3/2)}}\left[\psi\left(j+{3 \over 2}\right)-\psi(j+1)\right],
\eqno(2.28b)$$
where $2^j \Gamma(j+3/2)=\sqrt{\pi}(j+1/2)(2j-1)!!$, and
$$G={\pi \over 4}\ln 2+2\int_0^1 {{\sin^{-1}(y/\sqrt{2}) dy} \over
{(y+1)\sqrt{2-y^2}}}.  \eqno(2.28c)$$
For obtaining Eq. (2.28a) we apply Eqs. (18) and (21) of Ref. \cite{lunev}.
For Eq. (2.28b) we expand the integrand factor $1-x/2$ as geometric series,
writing
$$G=-{1 \over 4}\sum_{j=0}^\infty {1 \over 2^j}\int_0^1 x^j \ln\left({x \over 2}\right){{dx} \over \sqrt{1-x^2}}.  \eqno(2.29)$$
Performing the integral and using the series (\cite{hansen}, p. 151 or 174)
$$\sum_{j=0}^\infty {{j!} \over {2^j\Gamma(j+3/2)}} ={1 \over {\Gamma(3/2)}}\sum_{j=0}^\infty {{j!} \over {2^j (3/2)_j}}=\sqrt{\pi}, 
\eqno(2.30)$$ 
where $(a)_j$ is the Pochhammer symbol, gives Eq. (2.28b).  The sum in the
latter equation may be re-expressed by using the duplication formula of the
digamma function.  For Eq. (2.28c) we inserted an integral representation of
$\psi$ into Eq. (2.28b).

By using an integral representation
$$\mbox{coth}^{-1}z={1 \over z}\int_0^1 {{dt} \over {1-t^2/z^2}}, \eqno(2.31)$$
and interchanging integrations we obtain from Eq. (2.26) with $\theta$ as in 
Eq. (1.7)
$$\mbox{Cl}_2(\theta)=\cot^{-1}a \ln a +[\ln(a^2+1)-2\ln a]\tan^{-1} a 
-\int_0^1 {{\ln(1-t^2)} \over {1+a^2t^2}}dt, \eqno(2.32)$$ 
giving
$$G={\pi \over 4} \ln 2 - \int_0^1 {{\ln(1-t^2)} \over {1+t^2}}dt.  \eqno(2.33)$$
This last equation may also be obtained by combining two formulas in Ref. 
\cite{grad} (p. 555).
By expanding $(1+a^2t^2)^{-1}$ as a geometric series in Eq. (2.32) we obtain
the series representation 
$$\mbox{Cl}_2(\theta)=\cot^{-1}a \ln a +[\ln(a^2+1)-2\ln a]\tan^{-1} a 
-\sum_{j=0}^\infty (-1)^j {a^{2j} \over {(2j+1)}}\left[\psi\left(j+ {3 \over 2} \right)+\gamma \right]. \eqno(2.34)$$ 

We have also determined a BBP-type formula for the Catalan constant
\cite{bailey04}, as well as for a trilogarithmic constant.  We have
\newline{\bf Lemma 4}.  (a) We have the expression for $G=\mbox{Cl}_2(\pi/2)$
$$G={1 \over 4}\sum_{j=0}^\infty {1 \over {16^j}}\left[{4 \over {(8j+1)^2}}
-{2 \over {(8j+4)^2}}-{1 \over {(8j+5)^2}}-{1 \over {(8j+6)^2}}\right]
-{\pi^2 \over {32}}+{\pi \over 8}\ln 2,  \eqno(2.35)$$
(b)
$$\mbox{Re}~\mbox{Li}_3\left({{1+i} \over 2}\right)={1 \over {48}}\ln^3 2-{5 \over
{192}}\pi^2 \ln 2+{{35} \over {64}}\zeta(3), \eqno(2.36)$$
and (c)
$$8\sum_{j=0}^\infty {1 \over {16^j}}\left[{4 \over {(8j+1)^3}}
-{2 \over {(8j+4)^3}}-{1 \over {(8j+5)^3}}-{1 \over {(8j+6)^3}}\right]$$
$$=-{\pi^2 \over 2} \ln 2+14\zeta(3)+32 \mbox{Im} ~\mbox{Li}_3\left({{1+i} \over 2}\right).  \eqno(2.37)$$
Equation (2.35) is based upon the integral
$$-4\int_0^1 {{(y-1)\ln(y/\sqrt{2})dy} \over {y^4-2y^3+4y-4}}=G+{\pi^2 \over 32},
\eqno(2.38)$$
and the summatory relation
$$\sum_{j=0}^\infty {1 \over {16^j}}{1 \over {(8j+k)^2}}+{{\ln 2} \over 2}
\sum_{j=0}^\infty {1 \over {16^j}}{1 \over {(8j+k)}} =-2^{k/2}\int_0^{1/\sqrt{2}}
{x^{k-1} \over {1-x^8}} \ln x ~dx,  \eqno(2.39)$$
and turns out to be a combination of two known BBP-type formulas given in
Section 5 of Ref. \cite{bailey04}.  

Lemma 4 parts (b) and (c) are constructed starting from the relation
$$2^{k/2}\int_0^{1/\sqrt{2}} {x^{k-1} \over {1-x^8}}\ln^2 x ~dx
={1 \over 4}\sum_{j=0}^\infty {1 \over {16^j}}{1 \over {(8j+k)}}\left[\ln^2 2
+{{4 \ln 2} \over {8j+k}}+{8 \over {(8j+k)^2}}\right].  \eqno(2.40)$$
We then form the sums
$$8\sum_{j=0}^\infty {1 \over {16^j}}\left[{4 \over {(8j+1)^3}}
-{2 \over {(8j+4)^3}}-{1 \over {(8j+5)^3}}-{1 \over {(8j+6)^3}}\right]$$
$$+4 \ln 2\sum_{j=0}^\infty {1 \over {16^j}}\left[{4 \over {(8j+1)^2}}
-{2 \over {(8j+4)^2}}-{1 \over {(8j+5)^2}}-{1 \over {(8j+6)^2}}\right]$$
$$+\ln^2 2\sum_{j=0}^\infty {1 \over {16^j}}\left[{4 \over {(8j+1)}}
-{2 \over {(8j+4)}}-{1 \over {(8j+5)}}-{1 \over {(8j+6)}}\right]$$
$$=64\int_0^1 {{(y-1)\ln^2(y/\sqrt{2})dy} \over {y^4-2y^3+4y-4}}$$
$$=16G \ln 2-\pi \ln^2 2+{2 \over 3}i\ln^3 2-{5 \over 6}i\pi^2 \ln 2-32 i\mbox
{Li}_3\left({{1+i} \over 2}\right)+14\left(1+{5 \over 4}i\right) \zeta(3).
\eqno(2.41)$$
We then apply both the degree 2 binary BBP-type formula for $\pi$ and
part (a) of the Lemma, thereby eliminating the appearance of the constant
$16G\ln 2-\pi \ln^2 2$.  We then take the real and imaginary parts of the
resulting equation.  Since the imaginary part must vanish, we obtain 
part (b), and the real part gives the formula of part (c).

In regards to Eqs. (2.36) and (2.37), degree 3 binary BBP-type formulas are known for the constants $\zeta(3)$, $\ln^3 2$, $\pi \ln^2$, $\pi^2 \ln 2$, and $\pi^3$.
However Lemma 4(c) does not appear in the compendium \cite{bailey04}.  By means
of Landen's transformation for Li$_3$ or other functional relationships, the
value Li$_3[(1+i)/2]$ may be related to other trilogarithm function values.

Lemma 4(b) and (c) may be computationally useful for providing a spigot
algorithm for the constant Li$_3[(1+i)/2]$.  We note that we have the alternative
expressions
$$\mbox{Re}~\mbox{Li}_3\left({{1+i} \over 2}\right) =\sum_{n=1}^\infty \sum_{m=0}^n
\left[{n \choose {4m+4}} -{n \choose {4m+2}}\right]{1 \over {2^n n^3}},$$
$$\mbox{Im}~\mbox{Li}_3\left({{1+i} \over 2}\right) =\sum_{n=1}^\infty \sum_{m=0}^n
\left[{n \choose {4m+1}} -{n \choose {4m+3}}\right]{1 \over {2^n n^3}},$$
and that it may be possible to reach Eqs. (2.36) and (2.37) from them. 

We mention an extension of Lemma 4 for BBP-type formulas for other
polylogarithmic constants.  These may be developed by using the integrals
$$J_n={2^{n+1} \over n}\int_0^1 {{(y-1)\ln^n(y/\sqrt{2})dy} \over {y^4-2y^3+4y-4}}, \eqno(2.42)$$
and
$$\int_0^{1 \over \sqrt{2}} x^{k-1+8j} \ln^n x ~dx={{(-1)^n} \over {(8j+k)^{n+1}}}
\Gamma\left[n+1,{1 \over 2}(8j+k)\ln 2\right],  \eqno(2.43)$$
together with the property given in Eq. (3.8) for the incomplete Gamma function.
The evaluation of the integrals of Eq. (2.42) includes the constants
$G \ln^{n-1} 2$, $\pi^j \ln^k 2$, where $j+k=n+1$, Li$_n[(1\pm i)/2]$, $\ln^p 2 ~\mbox{Li}_q(1/2)$, with $p+q=n+1$, Li$_n[(1\pm i)/2]$, and $\ln^{n-2}2~ \zeta(3)$.

{\bf Remark} on sums of sine values.   Supplementary to Eq. (2.7) we have
the sum
$$\sin {{x \pi} \over {10}}+\sin {{3x \pi} \over {10}}+\sin {{7x \pi} \over {10}}
+\sin {{9x \pi} \over {10}}=\{\pm \sqrt{5},0\}, \eqno(2.44)$$
for $x=1,\ldots,20$, the $-$ sign holds on the right side for $x=11,\ldots,20$, and
the result is zero when $(x,10) \neq 1$, i.e., when $x$ is not co-prime with
$10$. 
Here and within the rest of this remark the notation on the right side of the
equation indicates that the expression on the left takes the values within the
set indicated.  Furthermore, we have the sine sums   
$$\sin {{x \pi} \over {12}}+\sin {{5x \pi} \over {12}}+\sin {{7x \pi} \over {12}}
+\sin {{11x \pi} \over {12}}=\{\pm \sqrt{6},0\}, \eqno(2.45)$$
for $x=1,\ldots,24$ and the result is zero when $(x,12) \neq 1$, 
$$\sin {{2x \pi} \over {11}}-\sin {{4x \pi} \over {11}}+\sin {{6x \pi} \over {11}}
+\sin {{8x \pi} \over {11}}+\sin {{10x \pi} \over {11}}=\left\{\pm {\sqrt{11} \over 2}, 0 \right \}, \eqno(2.46)$$
for $x=1,\ldots,22$ and the result is zero when $x \equiv 0$ (mod $11$), 
$$\sin {{2x \pi} \over {15}}+\sin {{4x \pi} \over {15}}+\sin {{8x \pi} \over {15}}
+\sin {{14x \pi} \over {15}}=\left\{\pm {\sqrt{15} \over 2}, 0\right\}, \eqno(2.47)$$
for $x=1,\ldots,15$ and the result is zero when $(x,15) \neq 1$,
$$\sin {{x \pi} \over {5}}+\sin {{2x \pi} \over {5}}+\sin {{3x \pi} \over {5}}
+\sin {{4x \pi} \over {5}}=\{\pm \sqrt{5 \pm 2 \sqrt{5}}, 0\}, \eqno(2.48)$$
for $x=1,\ldots,10$ and the result is zero when $x =2,4,5,6,8$, and $10$,
$$\sin {{x \pi} \over {5}}-\sin {{2x \pi} \over {5}}-\sin {{3x \pi} \over {5}}
+\sin {{4x \pi} \over {5}}=\{\mp \sqrt{5 \mp 2 \sqrt{5}}, 0\}, \eqno(2.49)$$
for $x=1,\ldots,10$ and the result is zero when $x =2,4,5,6,8$, and $10$, 
$$\sin {{x \pi} \over {8}}+\sin {{3x \pi} \over {8}}+\sin {{7x \pi} \over {8}}
=\left\{\pm{1 \over 2} \sqrt{10 \pm \sqrt{2}}, {1 \over \sqrt{2}},-1,0 \right\}, \eqno(2.50)$$
for $x=1,\ldots,16$ and the result is zero when $x \equiv 0$ (mod $8$),
and
$$\sin {{x \pi} \over {8}}+\sin {{5x \pi} \over {8}}+\sin {{7x \pi} \over {8}}
=\left\{\pm{1 \over 2}\sqrt{10 \pm \sqrt{2}}, -{1 \over \sqrt{2}},1,0 \right\}, \eqno(2.51)$$
for $x=1,\ldots,16$ and the result is zero when $x \equiv 0$ (mod $8$).  We omit
the values taken by a multitude of other sine combinations, including 
$\sin {{x \pi} \over {8}}\pm\sin {{7x \pi} \over {8}}$, $\sin {{x \pi} \over {8}}
\pm \sin {{5x \pi} \over {8}}$, and $\sin {{3x \pi} \over {8}}+\sin {{7x \pi} 
\over {8}}$.  Many of the combinations of sine values listed in this Remark
are related to factorizations of Chebyshev polynomials.
Such sums as given have implications for relations between Clausen function values.  We return to this topic in the Discussion section.



\pagebreak
\centerline{\bf Proof of Proposition 1}
\medskip

Rather than restrict attention to $I_7$, we consider the more general integrals
$$I(n) \equiv \int_{\pi/3}^{\pi/2} \ln^n \left|{{\tan t+\sqrt{7}} \over {\tan t-\sqrt{7}}} \right |dt=\int_{\sqrt{3}}^\infty \ln^n \left|{{u+\sqrt{7}} \over
{u-\sqrt{7}}}\right| {{du} \over {1+u^2}}. \eqno(3.1)$$
By splitting the integral and performing further changes of variable we have
$$I(n)=I^{(1)}(n)+I^{(2)}(n)=
\int_{\sqrt{3}}^{\sqrt{7}} \ln^n \left({{\sqrt{7}+u} \over
{\sqrt{7}-u}}\right) {{du} \over {1+u^2}}+
\int_{\sqrt{7}}^\infty \ln^n \left({{u+\sqrt{7}} \over
{u-\sqrt{7}}}\right) {{du} \over {1+u^2}} \eqno(3.2)$$
$$={\sqrt{7} \over 2}\left[\int_{r_{73}}^\infty {{\ln^n v ~dv} \over {(2v^2-3v+2)}}
+\int_1^\infty {{\ln^n v ~dv} \over {(2v^2+3v+2)}}\right].  \eqno(3.3)$$

By the use of Lemma 2(a), from the form given in Eq. (3.2), it is evident that
$$I^{(2)}(1)=\mbox{Cl}_2[\cos^{-1}(-3/4)]=\mbox{Cl}_2(\pi+\theta_-)=-\mbox{Cl}_2
(\pi+\theta_+).  \eqno(3.4)$$

We have
$${{I^{(1)}(n)} \over {2 \sqrt{7}}}={1 \over 4}\int_{r_{73}}^\infty {{\ln^n v ~dv} \over {(2v^2-3v+2)}}={1 \over 8}\int_{r_{73}}^\infty {{\ln^n v dv} \over {(v-v_+)
(v-v_-)}}$$
$$={1 \over {8(v_+-v_-)}}\int_{r_{73}}^\infty\left[{1 \over {v-v_+}}-{1 \over {v-v_-}} \right] \ln^n v ~dv, \eqno(3.5)$$
where $v_\pm=(3\pm i\sqrt{7})/4=\exp(i\theta_\pm)$.
By making the change of variable $y=1/v$ and expanding the bracketed terms in
geometric series, we have
$${{I^{(1)}(n)} \over {2 \sqrt{7}}}={{(-1)^n} \over {8(v_+-v_-)}}\int_0^{1/r_{73}} \ln^n y \sum_{\ell=0}^\infty \left({1 \over v_-^{\ell+1}}-
{1 \over v_+^{\ell+1}}\right) {{dy} \over y^{\ell+2}}.  \eqno(3.6)$$
We perform the integral in terms of the incomplete Gamma function $\Gamma(x,y)$,
finding that
$${{I^{(1)}(n)} \over {2 \sqrt{7}}}={{(-1)^{n+1}} \over {8(v_+-v_-)}}
\sum_{\ell=0}^\infty \left({1 \over v_-^{\ell+1}}-{1 \over v_+^{\ell+1}}\right)
{{\Gamma[n+1,-(\ell+1)\ln r_{73}]} \over {(\ell+1)^{n+1}}}.  \eqno(3.7)$$
We next use the property for $n \geq 0$ an integer 
$$\Gamma(n+1,x)=n!e^{-x}\sum_{m=0}^n {x^m \over {m!}}, \eqno(3.8)$$
to write
$${{I^{(1)}(n)} \over {2 \sqrt{7}}}={{(-1)^{n+1}n!} \over {8(v_+-v_-)}}
\sum_{\ell=1}^\infty \left({1 \over v_-^{\ell}}-{1 \over v_+^{\ell}}\right)
{r_{73}^\ell \over \ell^{n+1}} \sum_{m=0}^n {\ell^m \over {m!}} \ln^m \left(
{1 \over r_{73}}\right).  \eqno(3.9)$$
Since $v_+$ and $v_-$ are complex conjugates, the sum over $\ell$ in this
equation is pure imaginary, as is $v_+-v_-=i\sqrt{7}/2$.

We now specialize to $n=1$, whereby 
$${{I^{(1)}(1)} \over {2 \sqrt{7}}}={1 \over {8(v_+-v_-)}}\left [ 
\sum_{\ell=1}^\infty \left({1 \over v_-^{\ell}}-{1 \over v_+^{\ell}}\right)
{r_{73}^\ell \over \ell^2} -\ln r_{73}\sum_{\ell=1}^\infty \left({1 \over v_-^{\ell}}-{1 \over v_+^{\ell}}\right){r_{73}^\ell \over \ell} \right ]$$
$$={1 \over {8(v_+-v_-)}}\left [\mbox{Li}_2\left({r_{73} \over v_-}\right)
-\mbox{Li}_2\left({r_{73} \over v_+}\right) -\ln r_{73}\ln\left({{1-v_- r_{73}}
\over {1-v_+ r_{73}}}\right) \right], \eqno(3.10)$$
where we used the defining series form of Li$_2$, and the fact that $v_-v_+=1$.
The latter logarithmic value may be rewritten as
$$\ln\left({{1-v_- r_{73}} \over {1-v_+ r_{73}}}\right)=-2i\tan^{-1}
{1 \over 5}(2\sqrt{3} +\sqrt{7})=2i\omega_+, \eqno(3.11)$$
and we now use relation (1.15), taking advantage of the fact that the right
side of Eq. (3.10) is the ratio of two pure imaginary quantities.

We then find that
$${{I^{(1)}(1)} \over {2 \sqrt{7}}}={1 \over {4\sqrt{7}}}
[\mbox{Cl}_2(2\omega_+)-\mbox{Cl}_2(2\omega_++2\theta_+)+\mbox{Cl}_2(2\theta_+)],  \eqno(3.12)$$
where we repeatedly made use of $\omega_++\omega_-=0$.  We combine with the
result in Eq. (3.4) for $I^{(2)}(1)$ and use the duplication formula (1.17),
giving Eq. (1.13).

\centerline{\bf Discussion and additional results}
\medskip

Since our result for $I_7$ involves angles other than $\theta_7=2 \tan^{-1}\sqrt{7}$ of Eq. (26) of Ref. \cite{bb98}, another Clausen function value relation is suggested.  In fact, by Eqs. (4.61) and (4.63) of \cite{lewin} we have the relation 
$$2\mbox{Cl}_2(2\theta_+)-3\mbox{Cl}_2(2\theta_+-\theta_7)-\mbox{Cl}_2(3\theta_7
-2\theta_+)+6\mbox{Cl}_2(\pi+\theta_7)=0.  \eqno(4.1)$$
Since the duplication formula (1.17) holds, the value Cl$_2(\theta_+)$ is then
expressible as a combination of the Clausen function at other angles.
Additionally, as we have noted in Eq. (1.12b), we have $2\omega_+=\theta_7-4\pi/3$.

The sine relations given at the end of Section 2 lead to a series of other
Clausen function and hence Hurwitz zeta and polygamma function expressions.
For $2q$ even, let  
$$\mbox{Cl}_{2q}(\theta) \equiv \sum_{n=1}^\infty {{\sin n \theta} \over
n^{2q}}  \eqno(4.2)$$
be the generalized Clausen function.  Then, for instance, from Eq. (2.7) we 
have 
$$\mbox{Cl}_q\left({{2\pi} \over 7}\right)+\mbox{Cl}_q\left({{4\pi} \over 7}\right)-\mbox{Cl}_q\left({{6\pi} \over 7}\right)$$
$$={\sqrt{7} \over 2} 7^{-q}\left[\zeta\left(q,{1 \over 7}\right)+\zeta\left(
q,{2 \over 7}\right)-\zeta\left(q,{3 \over 7}\right)+\zeta\left(q,{4 \over 7}
\right)-\zeta\left(q,{5 \over 7}\right)-\zeta\left(q,{6 \over 7}\right)\right],
\eqno(4.3)$$
that presents an extension of Lemma 1.  We omit related expressions in terms
of the polygamma functions, as well as similar expressions coming from
additional sine function value relations.


The method of proof of Proposition 1 has furnished the more general result
for the integrals $I^{(1)}(n)$ in terms of polylogarithm functions Li$_j$ and
hence in terms of generalized Clausen functions.  From Eq. (3.9) we have
$${{I^{(1)}(n)}=(-1)^n}{i \over 2}\left \{\mbox{Li}_{n+1}\left({r_{73} \over v_-}\right)-\mbox{Li}_{n+1}\left({r_{73} \over v_+}\right) +
\mbox{Li}_{n}\left({r_{73} \over v_-}\right)-\mbox{Li}_{n}\left({r_{73} \over v_+}\right) +\ldots \right.$$
$$+{{\ln^j(1/r_{73})} \over {j!}}\left[\mbox{Li}_{n-j+1}\left({r_{73} \over v_-}\right)-\mbox{Li}_{n-j+1}\left({r_{73} \over v_+}\right) \right]+\ldots$$
$$\left.+{{\ln^{n-1}(1/r_{73})} \over {(n-1)!}}\left[\mbox{Li}_{2}\left({r_{73} \over v_-}\right)-\mbox{Li}_{2}\left({r_{73} \over v_+}\right) \right]\right\}
+(-1)^n {i \over 2} \ln^n (1/r_{73})\ln\left({{1-v_- r_{73}}
\over {1-v_+ r_{73}}}\right). \eqno(4.4)$$
Supplementing Eq. (4.2), one writes for $2r+1$ an odd integer 
$$\mbox{Cl}_{2r+1}(\theta) \equiv \sum_{n=1}^\infty {{\cos n \theta} \over
n^{2r+1}}.  \eqno(4.5)$$
Similarly, the integrals $I^{(2)}(n)$ and $I(n)$ may be evaluated.

Also following the method of Proposition 1 we may make the following extension
of Eq. (1.8).  Let $\theta_+=-\tan^{-1}(\sqrt{1-b^2}/b)$ and
$\omega_+=\tan^{-1}[\sqrt{1-b^2}/(a+b)]$.  Then we have
{\newline \bf Proposition 2}.  For real $a \geq 0$ and $|b|<1$ we have
$$I(a,b)\equiv \int_a^\infty {{\ln y ~dy} \over {y^2+2by+1}}
={1 \over 2}{1 \over \sqrt{1-b^2}}[\mbox{Cl}_2(2\omega_+)
-\mbox{Cl}_2(2\omega_++2\theta_+)+\mbox{Cl}_2(2\theta_+)] \eqno(4.6)$$
$$={1 \over 2}{1 \over \sqrt{1-b^2}}[\mbox{Cl}_2(2\theta_2-2\theta_1) -\mbox{Cl}_2(\pi-2\theta_1)+\mbox{Cl}_2(\pi-2\theta_2)], \eqno(4.7)$$
where
$\tan \theta_1=b/\sqrt{1-b^2}$ and $\tan \theta_2= (1/a+b)/\sqrt{1-b^2}$.

In obtaining Eq. (4.6), we write
$$I(a,b)=-{1 \over {(y_+-y_-)}}\int_0^{1/a}\ln v \left[{1 \over {(1-y_+v)}}
-{1 \over {(1-y_-v)}}\right]{{dv} \over v}, \eqno(4.8)$$
where $y_\pm \equiv -b \pm i \sqrt{1-b^2}=\exp(i \theta_\pm)$.  We then 
proceed similarly as in the proof of Proposition 1.  Of course Eq. (4.6)
may be rewritten with the aid of the duplication formula (1.17).

For Eq. (4.7), we integrate an equality with respect to $a$ and then
determine the constant of integration that may depend only upon $b$.
By the use of the chain rule, the relation
$${{d\theta_2} \over {da}}=-{\sqrt{1-b^2} \over {a^2+2ab+1}}, \eqno(4.9)$$
and the first expression on the right side of Eq. (1.14) for Cl$_2$, we easily
verify the equality of both sides of Eq. (4.7) upon differentiating with
respect to $a$.  The constant of integration may be found at $a=1$.
By Lemma 2(a) [cf. Eq. (2.11) with $n=1$] we have
$I(1,\pm b)=\mbox{Cl}_2(\cos^{-1} b)/\sqrt{1-b^2}$.  


Having determined the integral $I(a,b)$ by two different methods, we have
found a relation among Clausen function values.  
{\newline \bf Corollary 2}.  With $\theta_+$, $\omega_+$, $\theta_1$, and
$\theta_2$ as in Proposition 2, we have
$$\mbox{Cl}_2(2\omega_+)-\mbox{Cl}_2(2\omega_++2\theta_+)+ \mbox{Cl}_2(2\theta_+) =\mbox{Cl}_2(2\theta_2-2\theta_1) -\mbox{Cl}_2(\pi-2\theta_1) +\mbox{Cl}_2(\pi-2\theta_2). \eqno(4.10)$$
This relation seems to reflect the property $\tan^{-1}x+\cot^{-1}x=\tan^{-1}x
+\tan^{-1}(1/x)=\pi/2$ for $x >0$.

From Proposition 2 we recover many known special cases.  In particular,
we have $I(0,b)=0$, when $\omega_+=-\theta_+$.  In turn, we have
{\newline \bf Corollary 3}.  We have for $c>0$ and $0<t<\pi$
$$\int_0^\infty {{\ln x ~dx} \over {x^2+2xc \cos t +c^2}}={{\ln c} \over c}
{t \over {\sin t}},  \eqno(4.11)$$
recovering a formula of Ref. \cite{grad} (p. 533).  
In order to obtain Eq. (4.11) we use the scaling substitution $y=x/c$ 
and make use of the fact that $I(0,b)=0$.  

Due to the presence of the parameter $b$ in Proposition 2, we may obtain
expressions for integrals of the form
$$\int_a^\infty {{y^j \ln y ~dy} \over {(y^2+2by+1)^{j+1}}}={{(-1)^j} \over
{2^j j!}}{{\partial^j I(a,b)} \over {\partial b^j}}.  \eqno(4.12)$$

\bigskip
\centerline{\bf Acknowledgement}
This work was partially supported by Air Force contract number FA8750-06-1-0001.

\pagebreak

\end{document}